\documentclass{mem}
\usepackage{natbib}\usepackage{txfonts}\usepackage{balance}
\usepackage{graphicx}
\usepackage[a4paper,breaklinks,dvipdfm]{hyperref}
\idline{82}{1}
\begin{document}
\def\teff{$T\rm_{eff }$}
\def\kms{$\mathrm {km s}^{-1}$}

\title{Evolution of shocks and turbulence in the formation of galaxy clusters embedded in Megaparsec-scale filaments}

   \subtitle{}

\author{
S. \,Paul\inst{1,4},
L. \,Iapichino\inst{2},
F. \,Miniati\inst{3},
J. \,Bagchi\inst{1}
\and K. \,Mannheim\inst{4}}

  \offprints{S. Paul}

\institute{
Inter University Centre for Astronomy and Astrophysics, Post Bag 4, Pune University Campus, Pune-411007, India
\and
Zentrum f\"ur Astronomie der Universit\"at Heidelberg, Institut f\"ur Theoretische Astrophysik, Albert-Ueberle-Strasse 2,  D-69120 Heidelberg, Germany
\and
Physics Department, Wolfgang-Pauli-Strasse 27, ETH-Z\"urich, CH-8093 Z\"urich, Switzerland
\and
Institut f\"ur Theoretische Physik und Astrophysik, Universit\"at W\"urzburg, Am Hubland, D-97074 W\"urzburg, Germany \\
\email{surajit@iucaa.ernet.in}
}

\authorrunning{Paul et al.}

\titlerunning{Turbulence in merging galaxy clusters}

\abstract{Massive structures like cluster of galaxies, embedded in cosmic filaments, release enormous amount of energy through their interactions. These events are associated with production of Mpc-scale shocks and injection of considerable amount of turbulence, affecting the non-thermal energy budget of the ICM. In order to study this thoroughly, we performed a set of cosmological simulations using the hydrodynamical code Enzo. We studied the formation of clusters undergoing major mergers, the propagation of merger shocks and their interaction with the filamentary cosmic web. This interaction is shown to produce peripheral structures remarkably similar to giant radio relics observed, for example, in Abell 3376 and Abell 3667. We find a relatively long timescale (about 4 Gyr) for turbulence decay in the centre of major merging clusters. This timescale is substantially longer than typically assumed in the turbulent re-acceleration models, invoked for explaining the statistics of observed radio halos.
}
\maketitle{}

\section{Introduction}

Virialised cosmic structures are embedded
 in the web-like filamentary network, where clusters form at the junction of those
 filaments \citep{bond,doroc}. In the hierarchical structure formation framework,
  clusters are the largest objects in the universe that may have attained virialisation
   recently, and at present many clusters are still in the process of growing by accretion
   and mergers. 

In a major merger (the case when the mass ratio of the clumps approaches unity), the sudden increase in bulk motion of ICM produces Mpc-scale shock fronts
 and stirs the ICM creating large-scale turbulent eddies, with size up to several hundred kiloparsecs \citep{ricker}.
    Merger shocks  are thus particularly
   interesting for transforming thermal to kinetic energy, by
injection of volume-filling turbulence in the ICM \citep{sub}. Strong collisionless shocks are also capable of
producing high-energy cosmic-ray particles (CR) via diffusive shock acceleration
 mechanism (DSA; \citealt{blandf}). Turbulence
can also stochastically re-accelerate the ambient electrons \citep{brun3}
 and amplify magnetic fields by shock compression and dynamo action.

In the last years, numerical simulations of
idealised halo mergers were often performed 
(e.g., \citealt{heinz,asai,xiang}). Clearly, hydrodynamical
simulations of cluster evolution in a cosmological framework
are somehow complementary to the above cited approach, and
provide the final testbed for the idealised studies and the necessary
link to the observations.

In a recent work \citep{paul} we simulated several major merger events in a
non-artificial setup with the hydrodynamical code Enzo
\citep{oshea}. Our aim was to account for the shock propagation geometry,
the effect of the propagation of merger shocks and their role in injecting
turbulence in the ICM and especially in the cluster core. We present in the following the main results of that work.

\section{Simulations of cluster mergers}
\label{merging-setup}

Our simulations were performed with the Adaptive Mesh Refinement (AMR), grid-based hybrid (N-body plus hydrodynamical) code Enzo v.~1.0 \citep{oshea}. For the simulation details, we refer the reader to \citet{paul}.

We focus our study on merger events occurred between  $0.25 < z < 0.7$ and halos with mass $M > 10^{13}\ M_{\sun}$ at the time of merger. We chose only major mergers (merging mass ratio larger than 0.5). The final sample consists of seven mergers, spanning different mass ratios, merger redshifts, and total cluster masses \citep{paul}.

\subsection{Evolution of merger shocks}
\label{morphology}

\begin{figure}[]
\resizebox{\hsize}{!}{\includegraphics[clip=true]{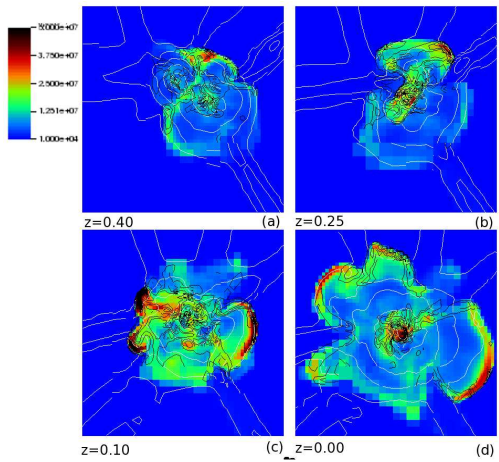}\\}
\caption{
\footnotesize
The evolution of a major merger is shown in slices. The redshift is indicated at the lower left of each panel, and an identification letter is at the lower right. Each panel has a size of $7.7 \times 7.7\ \mathrm{Mpc}\ h^{-1}$. The baryon density and vorticity are contoured in white and black respectively, and temperature is colour coded.
}
\label{vort_T_D}
\end{figure}

The morphological evolution of one of our clusters is shown in a map combining baryon density, temperature and vorticity (Fig.~\ref{vort_T_D}). In the case shown here, the two sub-clumps approach each other
with a relative velocity of $980\ \mathrm{km\ s^{-1}}$,
collide for the first time at $z \simeq 0.3$ and then pass through a core oscillation phase before getting relaxed around $z = 0$.  

The evolution of the baryonic component is the most prominent effect during a cluster merger. The gas in the ICM is severely attracted in the forming potential well, eventually generating a shock wave which propagates through the intra-cluster gas of the newly formed cluster. Gravitational and kinetic energy released in a merger event is thus dissipated into the ICM along with the evolution of shocks. This important feature of mergers is followed in Fig.~\ref{vort_T_D}. The temperature increase is first driven by compression at the centre of the forming cluster (Fig~\ref{vort_T_D}$a$), and subsequently the shock is launched and propagates outwards. 

The shape of the emerging shock depends on the mass of the merging clumps and on the geometry of the merger. In the case of Fig.~\ref{vort_T_D}, the shock front has a roughly ellipsoidal shape, with the arcs more pronounced along the merger axis. As this shock propagates out of the ICM of the newly formed cluster, it interacts with the surrounding filaments. The interaction with the web-like cosmic structure causes the breaking of the merger shock in separate sections, as clearly visible by comparing the temperature, density and vorticity shown in the bottom row of Figure~\ref{vort_T_D}. This interesting feature is obviously not modeled so far in simulations of idealised 
mergers (e.g.~\citealt{ricker,rick}), where symmetric bow-like shocks propagates unimpeded outwards.

\subsection{Injection and evolution of turbulence}

Velocity fluctuations are a distinctive feature of turbulent flows. It is therefore straightforward to relate the generation of turbulence with the vorticity of the flow, and to use this variable as a turbulence diagnostic in our analysis. 
In our reference run, vorticity is produced after the merger, just behind the shock, and propagates along with it. Some level of vorticity is also associated to both clumps before merging (Figure~\ref{vort_T_D}$a$) and also to the centre of the newly formed cluster (Figure~\ref{vort_T_D}$d$). A simple visual inspection of Figure~\ref{vort_T_D} thus suggest that major mergers stir the ICM effectively, resulting in a very volume-filling production of turbulence.

\begin{figure}[]
\resizebox{\hsize}{!}{\includegraphics[clip=true]{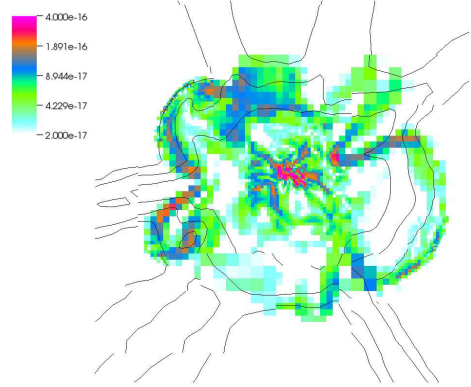}\\}
\caption{
\footnotesize
Slice of $7 \times 7\ \mathrm{Mpc}\ h^{-1}$, showing the vorticity $\omega^2$ for our reference cluster at $z = 0.05$. The vorticity is colour coded, whereas the density is represented as contours, in order to better highlight the filaments surrounding the cluster.
}
\label{vort_fila}
\end{figure}

An interesting by-product of shock propagation, highlighted in Figure~\ref{vort_fila}, is the interaction between merger shocks and filaments. It is already known that the warm-hot baryons flowing along the filaments results in a turbulent flow, when it mixes with the cluster gas \citep{nagai,maier}. Here, as an additional effect, we can see that a substantial level of vorticity is generated in the regions past the merger shock and surrounding filaments. In Figure~\ref{vort_fila} such `collars' can be clearly observed at the interaction surface of each filaments. The turbulence injection in these zones is probably related to the shearing (Kelvin-Helmholtz) instability between the shocked gas moving outwards and the filament gas flowing inwards. The level of vorticity is much larger than the turbulence associated with the baroclinic generation at the filament accretion shock. 

Observationally this feature can be related to the broken radio arcs (radio relics) seen in clusters like Abell 3376, Abell 3667 and CIZA-J2242 \citep{bagchi,hubb,van}. All these relics are associated with one common interesting feature i.e. the 'notch-like' structure, where the arc apparently bent inwards towards the cluster centre. It is natural, in the framework of the comparison with the simulations proposed above, to relate this observed morphology with the interaction of the merger shock with the filaments.

\subsection{Turbulence at the cluster centre}

\begin{figure}[]
\resizebox{\hsize}{!}{\includegraphics[clip=true]{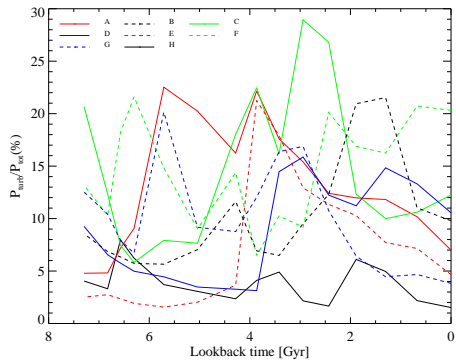}\\}
\caption{
\footnotesize
The time evolution of the turbulence pressure fraction in all studied clusters is here plotted. The black line is used for a relaxed reference cluster, the other ones indicate the merging clusters of our sample. From \citet{paul}.}
\label{turbu}
\end{figure}

As for the turbulence at the cluster core, in most of the clusters of our sample the turbulent pressure support is about $20 \%$ of the total pressure, after the merger. The ratio remains larger than $10 \%$ on a timescale of $2\ \mathrm{Gyr}$, and above the threshold of a nearly relaxed cluster (around $5 \%$) for about $4\ \mathrm{Gyr}$ (Fig.~\ref{turbu}). 
This timescale is a significant fraction of the cluster history, and much longer than the shock propagation timescale. A consequent and interesting question would pertain to any observable imprint that such event leaves on the cluster structure. A closely related issue concerns the bimodality in the correlation between radio and X-ray luminosity of clusters showing a radio halo, discussed by \citet{brun2} and \citet{brun}. In the light of the turbulent re-acceleration theory, this bimodality was interpreted as originating from the short timescale (of the order of $1\ \mathrm{Gyr}$) of the turbulence driving and decay. It is currently unclear how this acceleration scenario and the observed bimodality could be reconciled with the theoretical evidences of long turbulence decay timescales. 
A possible explanation is to relate the radio emission in halos only to the peak values of the turbulence evolution in cluster cores (cf.~\citealt{vazza}).

\begin{acknowledgements}
The computations were performed using the Enzo code, developed by the Laboratory for Computational Astrophysics at the University of California in San Diego (http://lca.ucsd.edu). The numerical simulations were carried out on the SGI Altix 4700 {\it HLRB2} of the Leibniz Computing Center in Garching (Germany). 
\end{acknowledgements}

\bibliographystyle{aa}

\end{document}